\documentclass[sigconf]{acmart}
\usepackage[caption=false,font=footnotesize]{subfig}
\usepackage{hyperref}
\usepackage{booktabs}
\usepackage{lipsum}
\usepackage{balance}
\usepackage{array,multirow,graphicx}
\usepackage{soul}
\usepackage[font=bf,skip=6pt]{caption}

\copyrightyear{2020} 
\acmYear{2020} 
\setcopyright{acmlicensed}
\acmConference[ETRA '20 Short Papers]{Symposium on Eye Tracking Research and Applications}{June 2--5, 2020}{Stuttgart, Germany}
\acmBooktitle{Symposium on Eye Tracking Research and Applications (ETRA '20 Short Papers), June 2--5, 2020, Stuttgart, Germany}
\acmPrice{15.00}
\acmDOI{10.1145/3379156.3391373}
\acmISBN{978-1-4503-7134-6/20/06}

\begin{document}

\title{Audiovisual Speech-In-Noise~(SIN) Performance of Young Adults with ADHD}

\author{Gavindya\,\,Jayawardena}
\affiliation{%
  \institution{Old Dominion University}
  \city{Norfolk}\state{VA}}
\email{gavindya@cs.odu.edu}
\author{Anne M. P. Michalek}
\affiliation{%
  \institution{Old Dominion University}
  \city{Norfolk}\state{VA}}
\email{aperrott@odu.edu}
\author{Andrew T. Duchowski}
\affiliation{%
  \institution{Clemson University}
  \city{Clemson}\state{SC}}
\email{duchowski@clemson.edu}
\author{Sampath Jayarathna}
\affiliation{%
  \institution{Old Dominion University}
  \city{Norfolk}\state{VA}}
\email{sampath@cs.odu.edu}

\begin{abstract}
Adolescents with Attention-deficit/hyperactivity disorder~(ADHD) have difficulty processing speech with background noise due to reduced inhibitory control and working memory capacity~(WMC).
This paper presents a pilot study of an audiovisual Speech-In-Noise~(SIN) task for young adults with ADHD compared to age-matched controls using eye-tracking measures.
The audiovisual SIN task consists of varying six levels of background babble, accompanied by visual cues.
A significant difference between ADHD and neurotypical (NT) groups was observed at 15 dB signal-to-noise ratio (SNR).
These results contribute to the literature of young adults with ADHD.

\end{abstract}



\maketitle

\section{Introduction}

The recent estimated prevalence of diagnosed ADHD in children and adolescents has increased from 6.1\% to 10.2\% over the period of 1997 to 2016 in the U.S. \cite{xu2018twenty}. 
Adolescents with ADHD have difficulty meeting time limits, controlling anger, inhibiting responses, and processing auditory information \cite{fields2017adult,fostick2017effect,barkley1997behavioral}.
Processing speech in background noise requires fundamental language abilities, higher working memory, as well as a higher signal-to-noise ratio (SNR) \cite{schneider2007competing}.
Since a person's ability to process speech with background noise depends on that person's auditory and cognitive system \cite{schneider2007competing}, young adults with ADHD may experience difficulty processing auditory information in the presence of background noise due to reduced inhibitory control \cite{barkley1997behavioral,woltering2013neurophysiological,woods2002neuropsychological,pazvantouglu2012neuropsychological}, and decreased working memory capacity (WMC) \cite{banich2009neural, alderson2013attention,michalek2014effects}. 

Unlike noise, which degrades listening conditions, the presence of external visual cues such as written, contextual information and facial movements, can enhance the processing of auditory information, especially when accompanied by noise \cite{jaaskelainen2010role,van2005visual,von2008simulation,rudner2009cognition,mishra2013seeing,fraser2010evaluating,michalek2014effects,moradi2013gated}. At increased noise levels, semantically related visual cues have a positive impact on the perception of spoken sentences \cite{zekveld2011influence}.
When increased noise is present during face-to-face conversation, adults tend to fixate more on the nose and mouth area of the speaker \cite{buchan2008effect}, confirming that oral-motor movements of the speaker aides speech recognition \cite{bristow2008hearing}.

Neurotypical (NT) individuals are known to perceive audiovisual cues more accurately from the right visual field (RVF)
than from the left visual field (LVF) \cite{kimura1973asymmetry}.
Multiple studies on this \cite{carter1995asymmetrical,voeller1988attention,mitchell1990reaction,heilman1991possible} showed the presence of a lateralized deficit in the visual-spatial attention of ADHD subjects,
which orients their attention to LVF targets.

Our work presents the performance of young adults with ADHD compared to age-matched controls using eye-tracking measures during an audiovisual SIN task. Our findings are consistent with the possibility that audiovisual cues, in general, are processed in such a way that WMC or cognitive load are not consistently impacted in increasing levels of background noise for NT adults \cite{michalek2018independence}.

\section{Methodology}

\subsection{Participants}
Our pilot study consisted of five young adults (4 F, 1 M) with a prior diagnosis of ADHD, and six NT young adults (4 F, 2 M) as the control. All participants were aged between 18 - 30 years, with no history of psychotic symptoms and normal
vision. 
Participants with a diagnosis of ADHD confirmed 
their diagnosis through medical documentation, including records from a physician or licensed psychiatrist. 
They were asked to remain medication-free for 12 hours prior to study participation. There were no participants who had been prescribed long lasting non-stimulants, so the 12-hour time frame was sufficient for all participants.
Information on the risks of avoiding medication were provided prior to the experiment, and participants acknowledged 
it
by signing 
a consent form approved by University's Institutional Review Board.
Both ADHD and NT participants went through a hearing screening of 20 dB HL at frequencies 500 Hz, 1000 Hz, 2000 Hz, and 4000 Hz, bilaterally, to ensure their hearing was within normal limits.

\begin{figure*}[hbt!]
	\centering
	\subfloat[Speaker]{%
		\label{fig:speaker}
		\includegraphics[width=.2\textwidth,height=2.7cm]{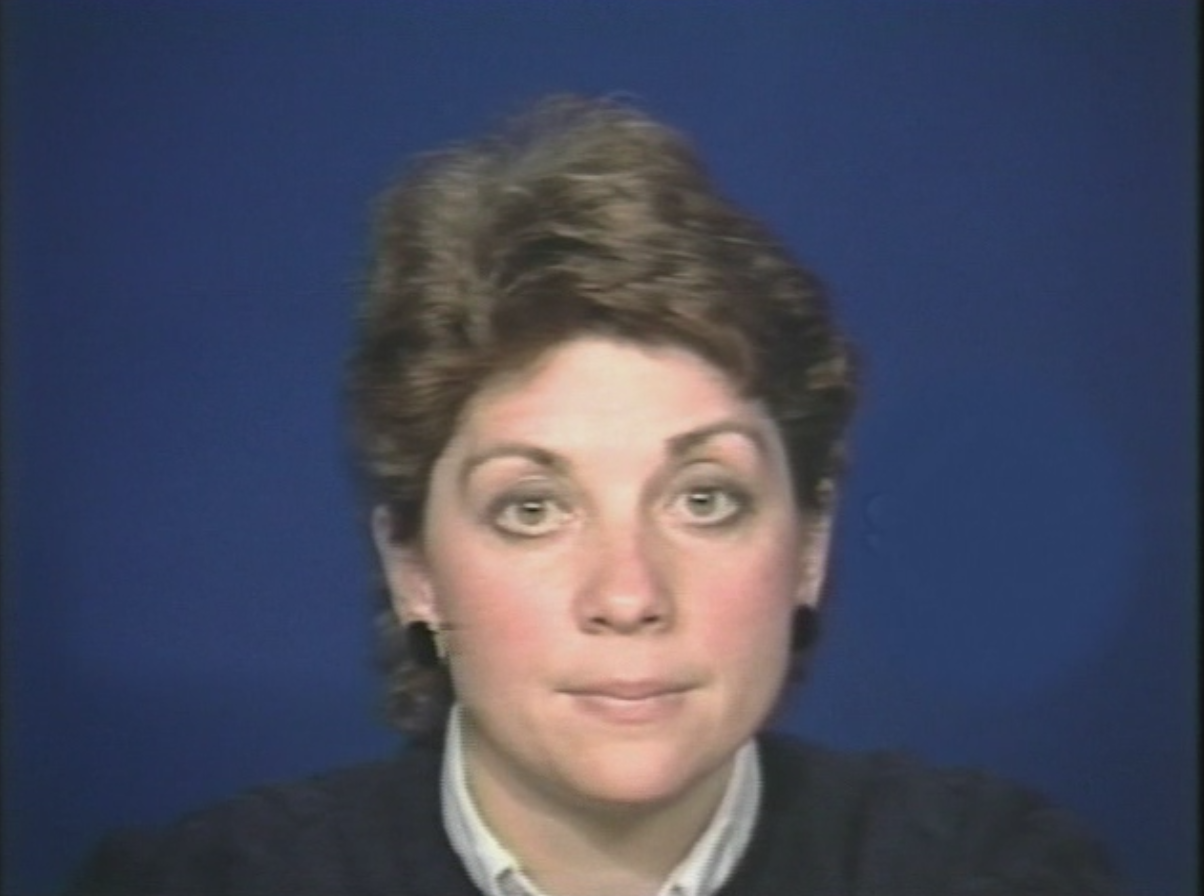}
	}
	\quad
	\subfloat[AOIs]{%
		\label{fig:aoi}
		\includegraphics[width=.2\textwidth,height=2.7cm]{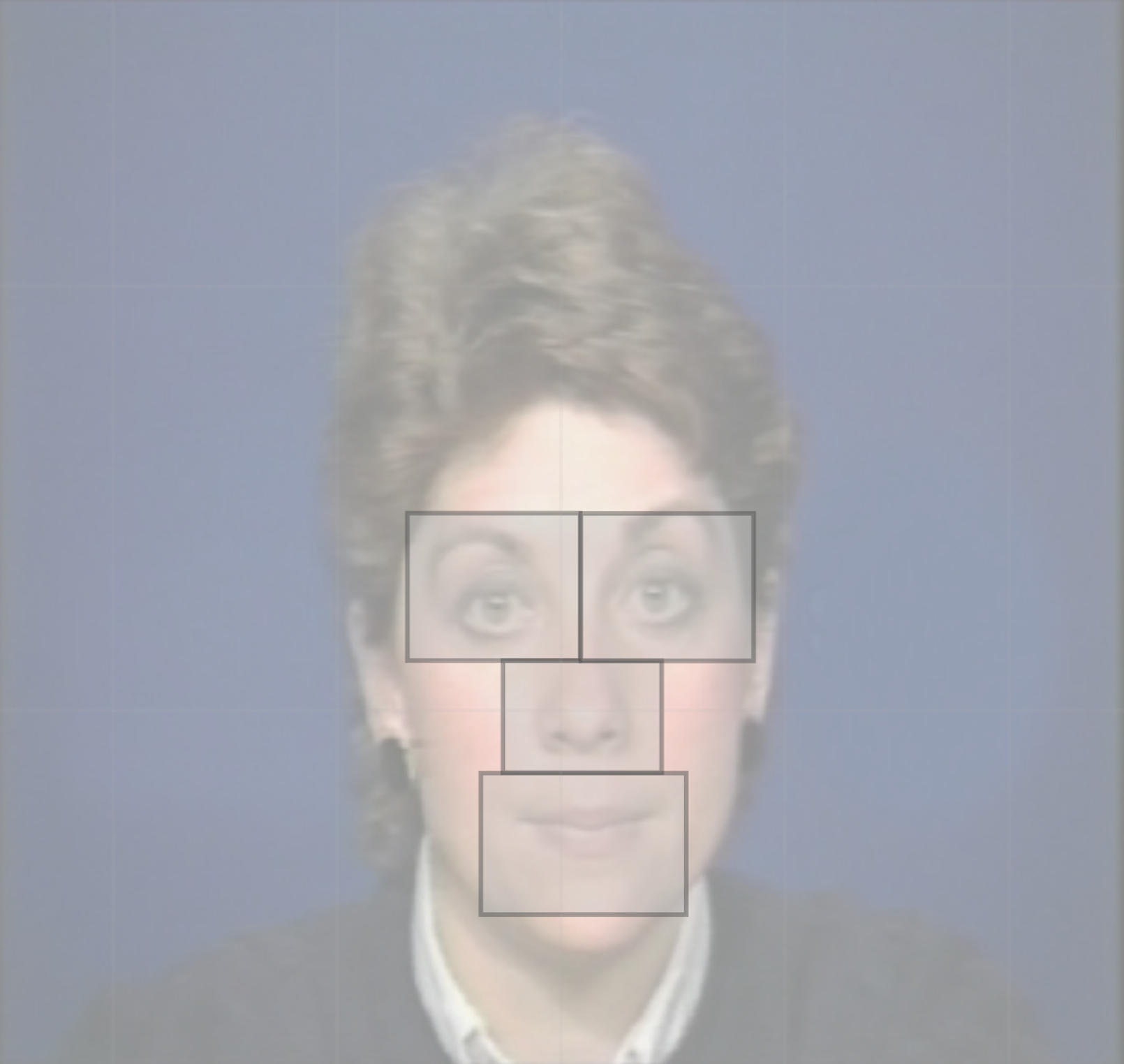}
	}
	\quad
	\subfloat[Scanpath]{%
		\label{fig:fxtn}
		\includegraphics[width=.2\textwidth,height=2.7cm]{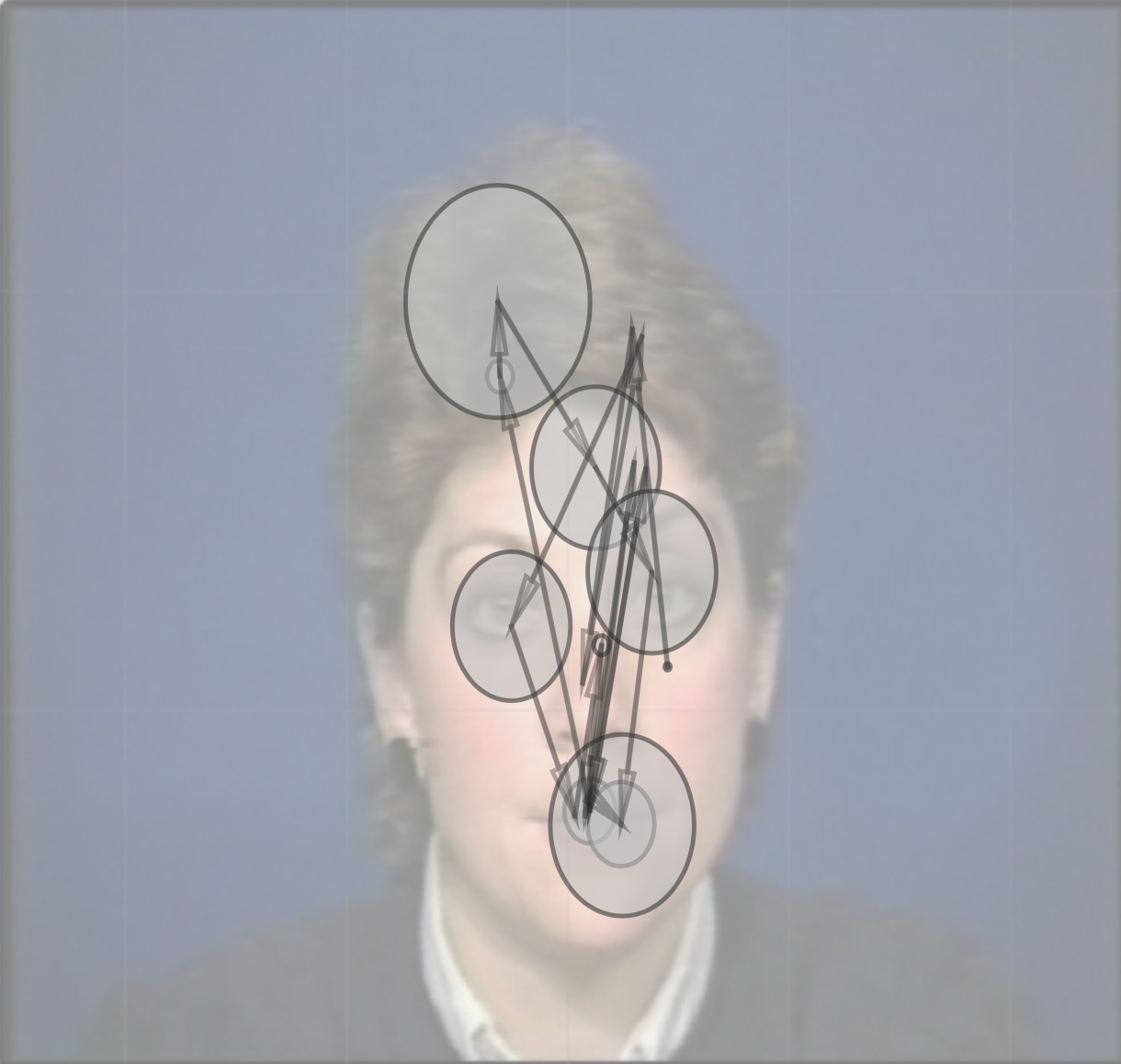}
	}
	\quad
	\subfloat[Fixations on AOIs]{%
		\label{fig:aoi_fxtn}
		\includegraphics[width=.2\textwidth,height=2.7cm]{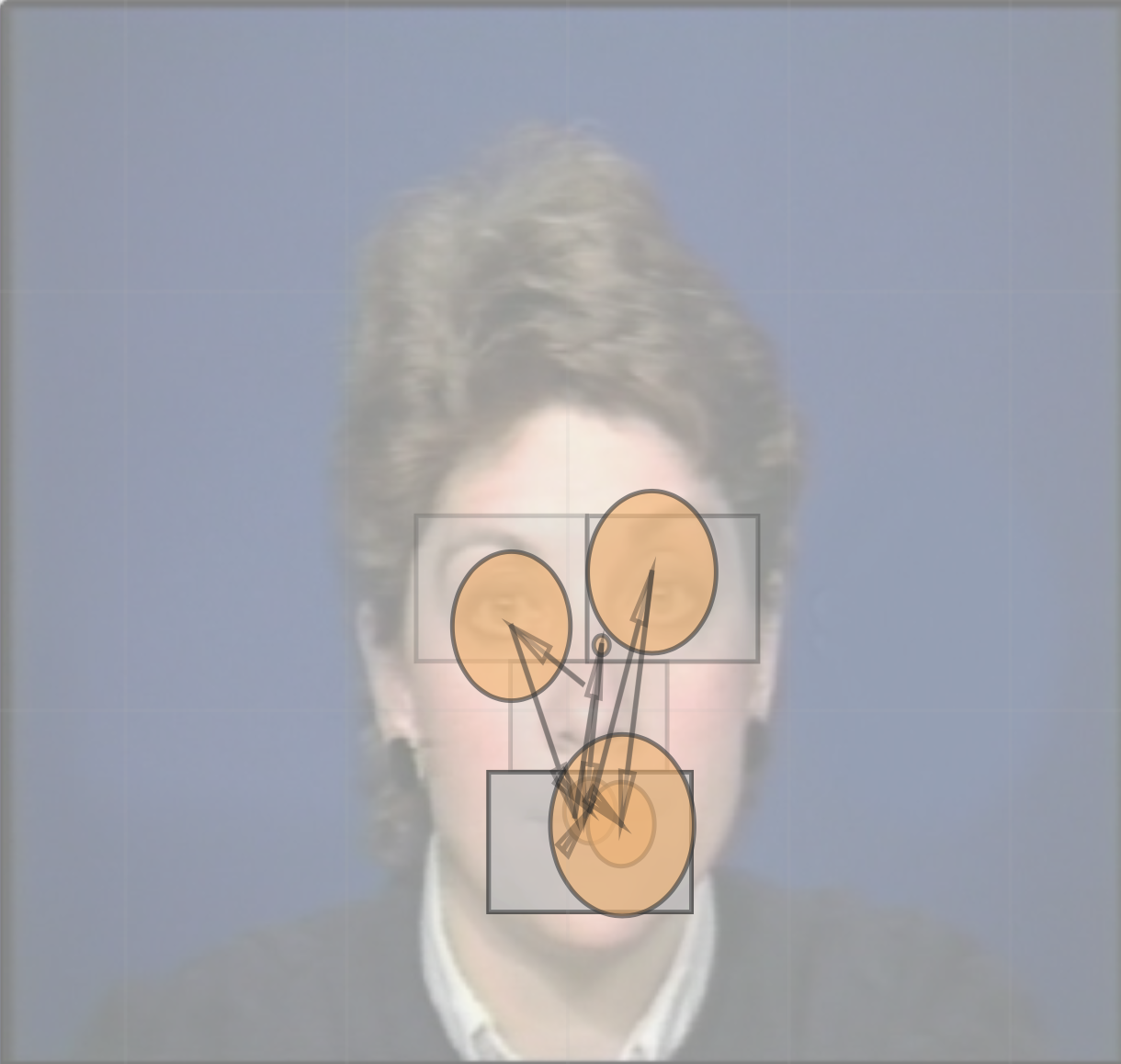}
	}
	\caption{ The Audiovisual QuickSIN Setup,
	\textmd{
	\protect\subref{fig:speaker} Speaker's face as viewed by the participants during the audiovisual SIN task,
	\protect\subref{fig:aoi} Four AOIs created for the eye-movement analysis: left eye, right eye, nose, and mouth, 
	\protect\subref{fig:fxtn} Sample scan-path with fixations , and \protect\subref{fig:aoi_fxtn} fixations on the AOIs of a participant while listening to one sentence.}}
	\label{fig:face}
\end{figure*}

\subsection{Speech-in-Noise Task} 

We used QuickSIN {\cite{killion2004development}} software to simultaneously present a 
sentence repetition task with background noise (i.e. speech babble) at six SNRs: 25 dB, 20 dB, 15 dB, 10 dB, 5 dB, and 0 dB. Each SNR represents the ratio of the dB level of speech to dB level of noise.
The level of background noise increases as the SNR decreases.
The audiovisual QuickSIN setup is presented in Figure \ref{fig:speaker}.
Participants were asked to listen to the sentences while simultaneously viewing the speaker's face
and then repeat each sentence verbally.
Participants were presented with nine sentence sets, each having six sentences representing all background noise levels.
Each sentence had an average of 8-13 words including five keywords
(e.g., \textit{The \underline{weight} of the \underline{package} was \underline{seen} on the \underline{high} \underline{scale}}).
Participants were scored based on the number of keywords accurately
repeated
per sentence.
The presentation of the nine sentence blocks was randomized and counterbalanced across participants.

\subsection{Eye-tracking Setup}

We used Tobii Pro X2-60 computer screen-based eye tracker (60 Hz, 0.4$^{\circ}$ accuracy) to record the eye movements of participants during the QuickSIN task.
Prior to the experiment, each participant was calibrated using Tobii's standard calibration methods.
We used Tobii Studio analysis software to pre-process gaze metrics using the I-VT filter (velocity threshold set to 30$^{\circ}$/second)
to extract eye movement metrics recorded throughout the study.

We specified four areas of interest (AOIs): 1) left eye, 2) right eye, 3) nose, and 4) mouth of the eye-tracking stimulus
to analyze the eye-movements of participants (see Figure \ref{fig:aoi}).

\subsection{Analysis}

To observe how the eye-tracking measurements change with audiovisual cues, we used our RAEMAP~\cite{jayawardena2020} eye movement processing pipeline, which is a modified version of gaze analytics pipeline~\cite{duchowski2017gaze}.
Upon correct mapping of variables, the original gaze analytics pipeline has the capability of extracting raw gaze data from various eye trackers \cite{duchowski2017gaze}.
After extracting raw gaze data, the gaze analytics pipeline: 
(1) classify raw gaze points into fixations, and 
(2) aggregate fixations related information for statistical analysis.
The gaze analytics pipeline facilitates computation of numerous eye movement metrics.
Also, it has the capability of generating visualizations of gaze points, fixations within AOIs, heat maps, ambient/focal fixations, and microsaccades per scan path.
The current implementation of the gaze analytics pipeline handles eye-tracking data recorded during each task of each person sequentially.
This process is computationally expensive, where the split and merge approach generates large number of intermediate files along the way of eye gaze metrics calculations.

\begin{figure*}[h]
  \centering
  \includegraphics[width=.9\textwidth]{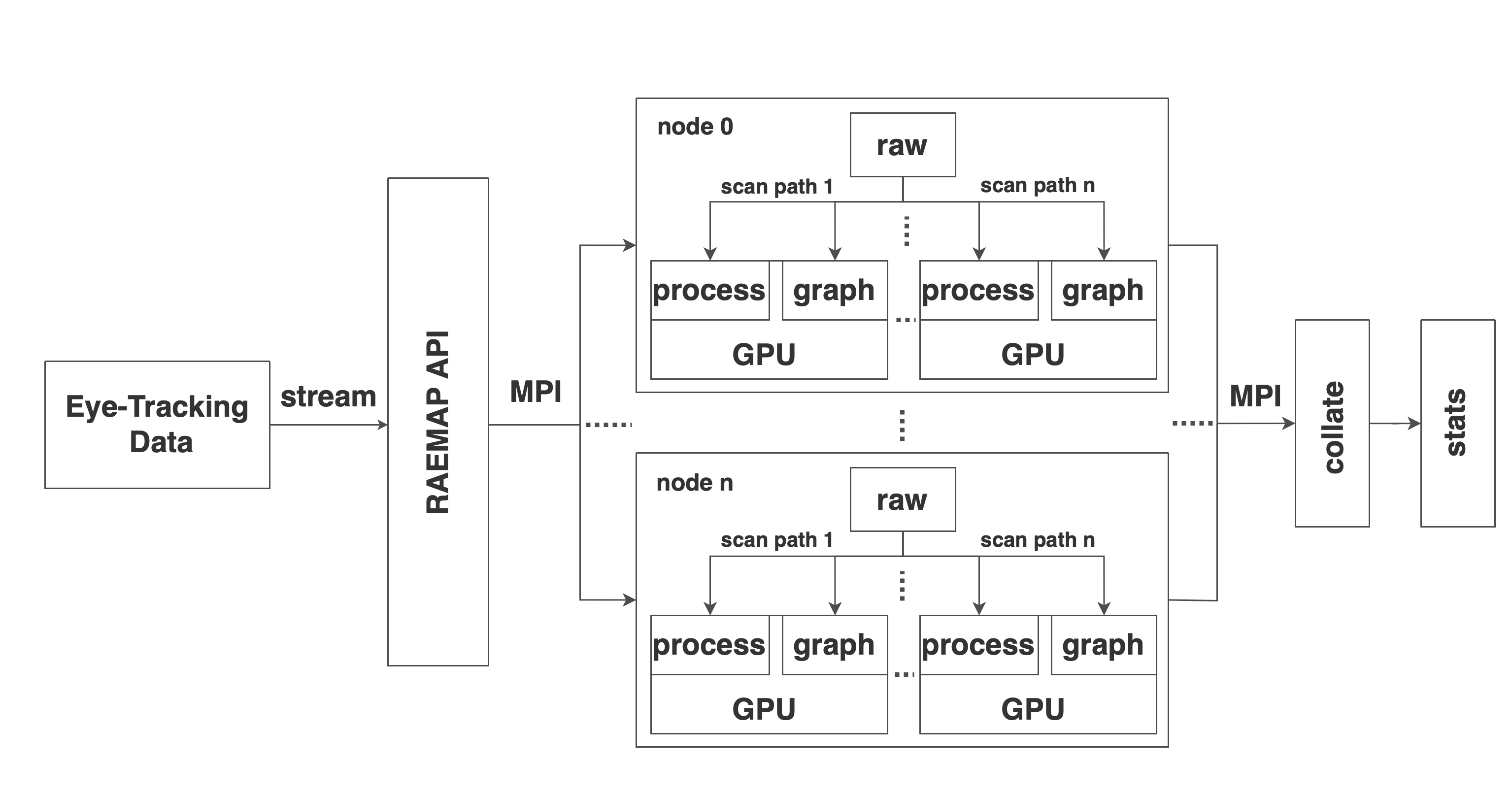}
  \caption{The architecture of RAEMAP which could process eye-tracking data as being streamed by an eye-tracker.
    \textmd{
    RAEMAP API distributes tasks among the nodes using MPI. 
    Each node hosts an instance of the RAEMAP providing the functionality \textit{raw} to extract raw gaze data, along with parallel processing of process and graph steps. \textit{Process} step calculate fixations, fixations in AOIs, saccade amplitudes, saccade duration, and IPA, whereas \textit{graph} step generate visualizations.
    MPI gather function facilitates the aggregation of calculated eye gaze metrics in \textit{collate} step, which provides data for statistical analysis in \textit{stats} step.
    }
    }
  \label{fig:raemap}
\end{figure*}

RAEMAP is developed such that calculations of eye gaze metrics utilize distributed computing resources as illustrated in Figure \ref{fig:raemap}.  
RAEMAP facilitates computation of traditional positional gaze metrics such as fixation count and fixation duration, as well as advanced metrics such as 
gaze transition entropy \cite{krejtz2015gaze}, and complex pupillometry measurements such as index of pupillary activity (IPA) \cite{duchowski2018index} which indicate cognitive load. 
RAEMAP also has the capability of generating visualizations of gaze points, AOIs, scan paths, and fixations on AOIs (see Figure \ref{fig:face}).
The architecture of RAEMAP is shown in Figure~\ref{fig:raemap}.

In RAEMAP, the calculations of eye gaze metrics of subjects are done in separate processes utilizing distributed computing resources as illustrated in Figure \ref{fig:raemap} since they are independent of one another to enhance the efficiency. 
The aggregation of calculated eye gaze metrics of all participants in each task is done using Message Passing Interface (MPI).
In addition, RAEMAP have the stream processing capability to calculate eye gaze metrics and visualize the scan path as data is being streamed by the eye tracker.

We applied RAEMAP to calculate gaze points, AOIs, scan paths, and fixations on AOIs per each sentence of the QuickSIN task for each participant. 
Figure \ref{fig:fxtn} shows a visualization of fixations of one participant while watching one sentence in the QuickSIN task.
Figure \ref{fig:aoi_fxtn} shows a visualization of fixations on pre-defined AOIs.
We generated gaze transition matrices and corresponding gaze transition entropies for both participant groups.
We also calculated the IPA counts for participants in both groups.

\section{Results}

We first report the performance of ADHD and NT participants during the QuickSIN task. Next we analyze changes in eye movements in relation to the six SNRs. A mixed, repeated measures ANOVA using a 2x6 design with main factors of group (ADHD or NT) and SNR (0 dB to 25 dB with 5dB increments) was carried out on the performance of QuickSIN task and the eye-tracking measures.

\subsection{QuickSIN performance}

We first analyze the performance of both ADHD and NT participants at each SNR. 
Each participant was assigned a score for every sentence,
based on the number of keywords accurately repeated out of five.
There was no main effect of the participant group for QuickSIN performance $F(1,9)=1.97, p>0.05$, 
indicating that performance was similar between ADHD and NT participants.
There was a significant main effect of the SNR on QuickSIN performance, $F(1.23,11.11)=127.78, p<0.001$, 
indicating that performance was different among SNRs.
There was no significant interaction effect between SNR and participant group, $F(1.23,11.11)<1,p>0.05$.
To further evaluate the main effect of the SNR, we conducted a $t$-test for each SNR, identifying a significant difference of QuickSIN performance between the two groups at 15 dB SNR, $p<0.05$.

The performance of NT participants was best at 15 dB SNR whereas, the performance of participants with ADHD was best at 20 dB SNR.
In general, when the task's difficulty level was easy (SNR>15 dB), both ADHD and NT participants performed well by recalling 4.7 keywords out of 5 on average per sentence.
In contrast, when the task was difficult (0 dB SNR), both ADHD and NT participants did not perform well by recalling 2.3 keywords out of 5 on average per sentence.
At 15 dB SNR, participants with ADHD recalled 4.7 keywords on average and NT participants recalled 4.9 keywords on average.

\subsection{Analysis of Fixation Count}

Fixation count indicates the number of times eyes fixated on an AOI. 
We observed that participants with ADHD fixate more on left eye whereas NT participants fixate more on right eye.
At SNRs 20 dB, 15 dB, and 10 dB, participants with ADHD fixated mostly on the left eye region.

\begin{table*}[hbt!]
    \centering
    \caption{Fixation counts on AOIs of ADHD and NT Participants.}
    \label{tab:fixation_counts}
    \begin{tabular}{|r|c|c|c|c|c|c|c|c|}
        \hline
        \multirow{2}{*}{SNR} & \multicolumn{2}{c|}{Left Eye} & \multicolumn{2}{c|}{Right Eye} & \multicolumn{2}{c|}{Nose} & \multicolumn{2}{c|}{Mouth} \\
        \cline{2-9}
                & ADHD           & NT            & ADHD          & NT            & ADHD          & NT            & ADHD          & NT            \\
        \hline
        25 dB &    $ 93.6\pm14.5$    & $54.0\pm13.2$    & $93.4\pm23.1$    & $125.7\pm21.1$    & $94.4\pm13.5$    & $44.3\pm12.3$    & $115.6\pm23.7$    & $150.0\pm21.7$ \\
        
        20 dB &    $105.4\pm10.9$    & $52.3\pm10.0$    & $74.0\pm18.7$    & $ 96.7\pm17.1$    & $85.4\pm11.4$    & $51.7\pm10.4$    & $ 71.0\pm19.2$    & $116.5\pm17.5$ \\
        
        15 dB &    $ 83.6\pm9.20$    & $68.7\pm8.40$    & $60.6\pm9.70$    & $ 76.2\pm8.80$    & $68.4\pm12.6$    & $48.5\pm11.5$    & $ 76.0\pm15.3$    & $ 85.2\pm13.9$ \\
        
        10 dB &    $ 60.2\pm6.10$    & $42.3\pm5.60$    & $42.2\pm3.70$    & $ 51.5\pm3.40$    & $52.4\pm10.0$    & $44.7\pm9.10$    & $ 46.4\pm8.80$    & $ 66.7\pm8.10$ \\
        
         5 dB &    $ 33.8\pm3.90$    & $30.0\pm3.60$    & $34.2\pm2.70$    & $ 27.3\pm2.40$    & $42.6\pm12.9$    & $29.5\pm11.8$    & $ 36.2\pm4.50$    & $ 35.7\pm4.10$ \\
         
         0 dB &    $ 13.8\pm5.80 $    & $10.2\pm5.30 $    & $19.0\pm6.90 $    & $  5.30\pm6.30 $    & $41.6\pm16.9$    & $34.8\pm15.5$    & $ 18.8\pm6.80 $    & $ 16.0\pm6.20 $ \\
        \hline
    \end{tabular}
    \vspace{-2pt}
\end{table*}

We conducted repeated measures 2x6 two-way ANOVA with main factors of group, and SNR on fixation counts on each AOI.
We observed a significant main effect of the SNR, $F(2.5,22.5)=22.14, p<0.001$ as well as group, $F(1,9)=12.27, p<0.008$ on fixation counts on left eye indicating that number of fixations differed among ADHD and NT participants as well as different SNRs.
There was a significant interaction effect between SNR and group, $F(2.5,22.5)=2.958, p<0.05$, indicating that fixation counts on the left eye on different listening conditions differed depending on the ADHD diagnosis.

We observed a significant main effect of the SNR for fixations on right eye, nose, and mouth, 
all $p<0.02$, but no main effect of the group, 
all $F(1,9)<2.6, p>0.05$.
Also, there was no significant interaction effect between SNR and group for right eye, nose, and mouth, 
all $p>0.05$.
Contrasts of the SNR revealed that the number of fixations on the nose significantly differed when compared 25 dB, 20 dB, and 15 dB SNRs against 0 dB, all $F(1,9)>5.3, p<0.05$ among the two groups.
The number of fixations on left eye, right eye, and mouth on all SNRs significantly differed when compared 0 dB, $p<0.05$ (see Table \ref{tab:fixation_counts}).

\subsection{Gaze Transition Matrices}

\begin{figure*}[hbt!]
	\centering
	\subfloat[ADHD:25 dB]{%
		\label{fig:adhd_25}
		\includegraphics[width=.14\textwidth,trim={0 0 0 50},clip]%
		{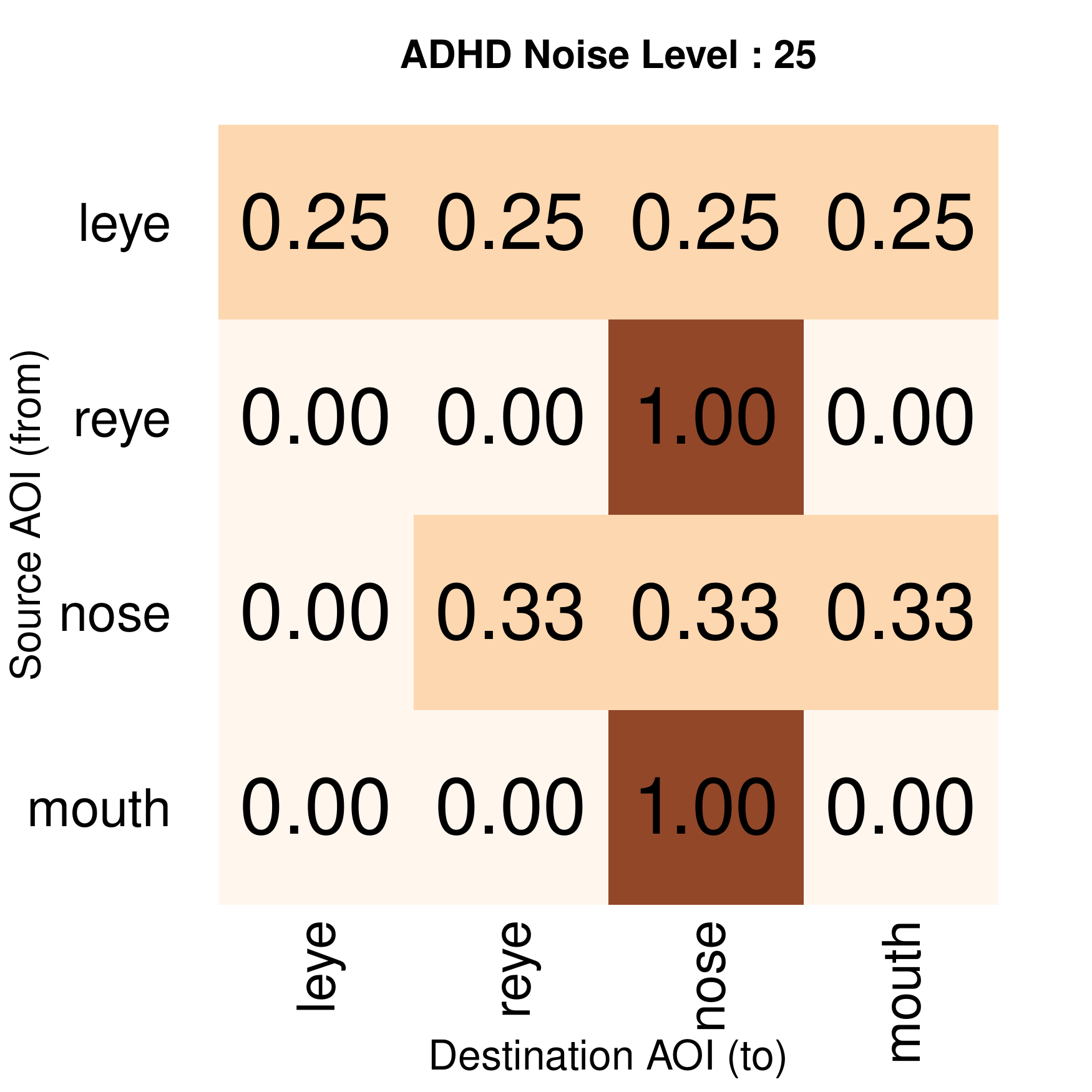}
	}
	\quad
	\subfloat[ADHD:20 dB]{%
		\label{fig:adhd_20}
		\includegraphics[width=.14\textwidth,trim={0 0 0 50},clip]%
		{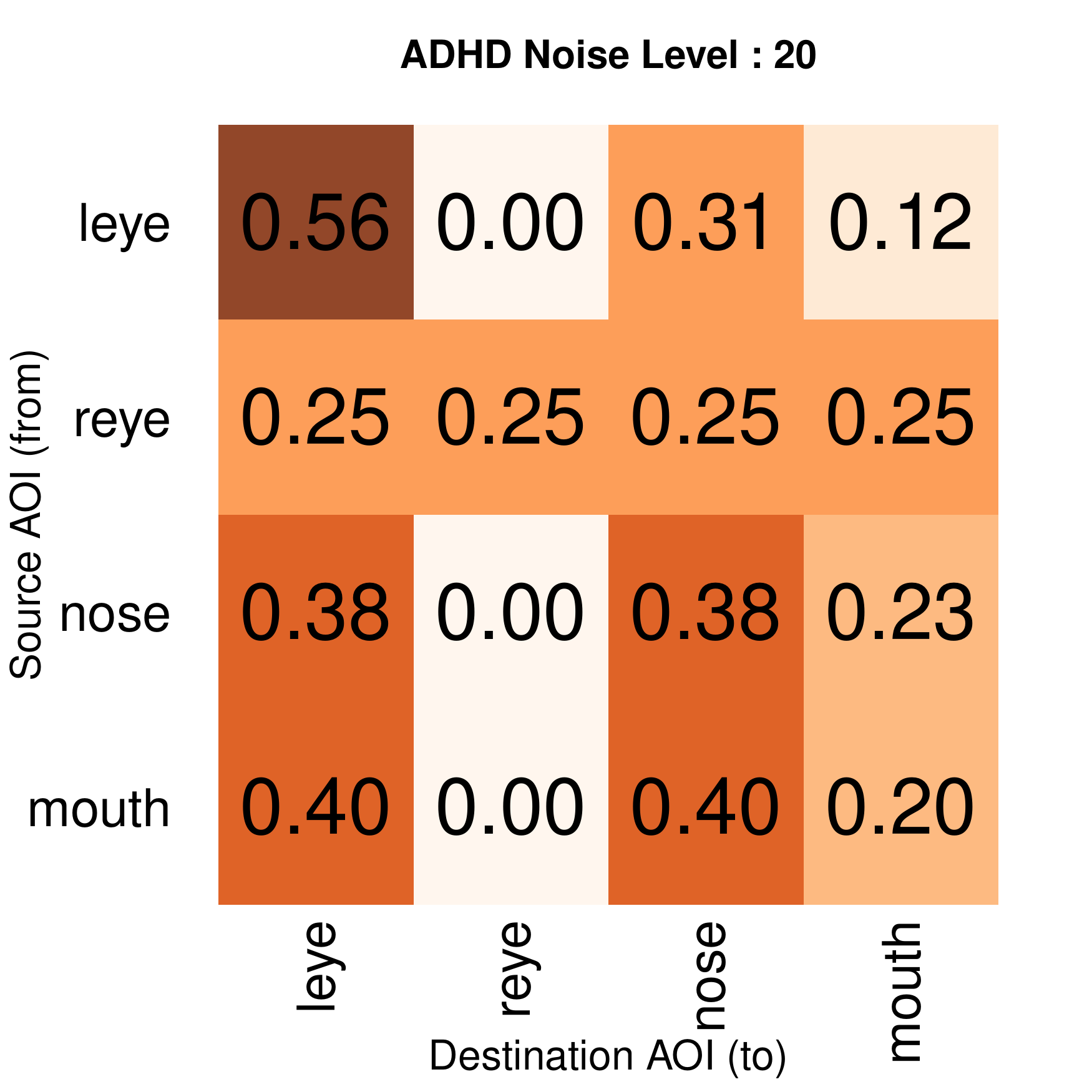}
	}
	\quad
	\subfloat[ADHD:15 dB]{%
		\label{fig:adhd_15}
		\includegraphics[width=.14\textwidth,trim={0 0 0 50},clip]%
		{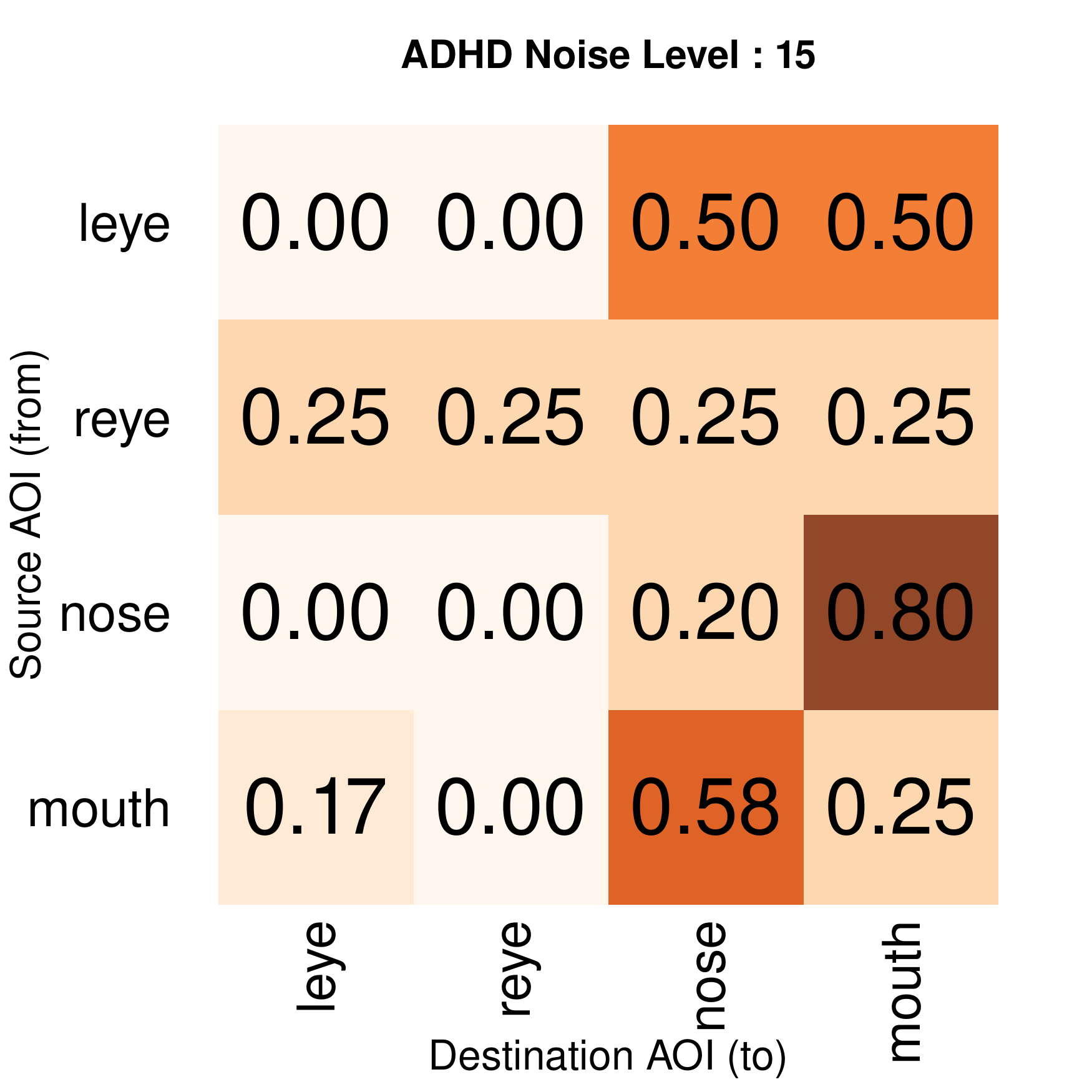}
	}
	\quad
	\subfloat[ADHD:10 dB]{%
		\label{fig:adhd_10}
		\includegraphics[width=.14\textwidth,trim={0 0 0 50},clip]%
		{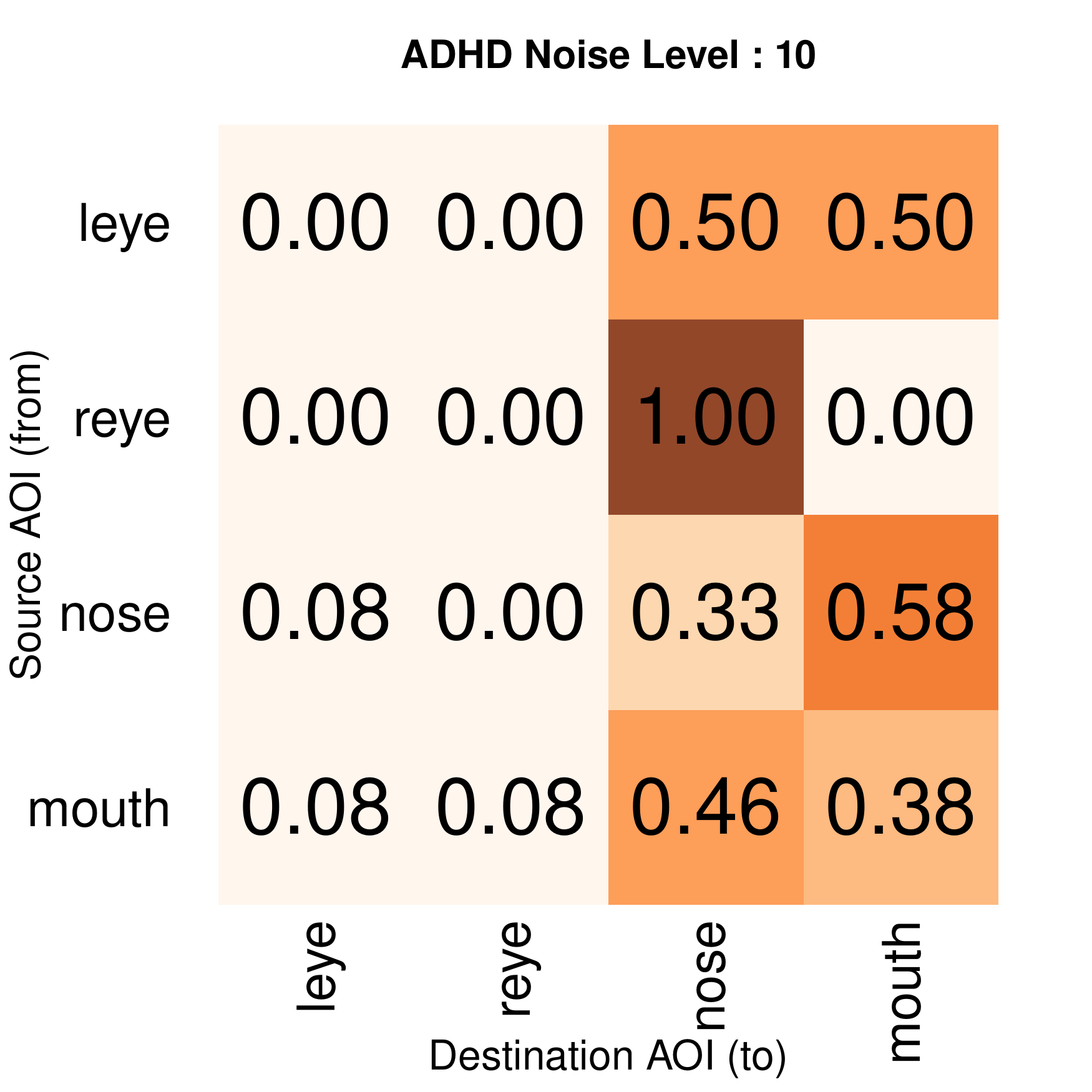}
	}
	\quad
	\subfloat[ADHD:5 dB]{%
		\label{fig:adhd_5}
		\includegraphics[width=.14\textwidth,trim={0 0 0 50},clip]{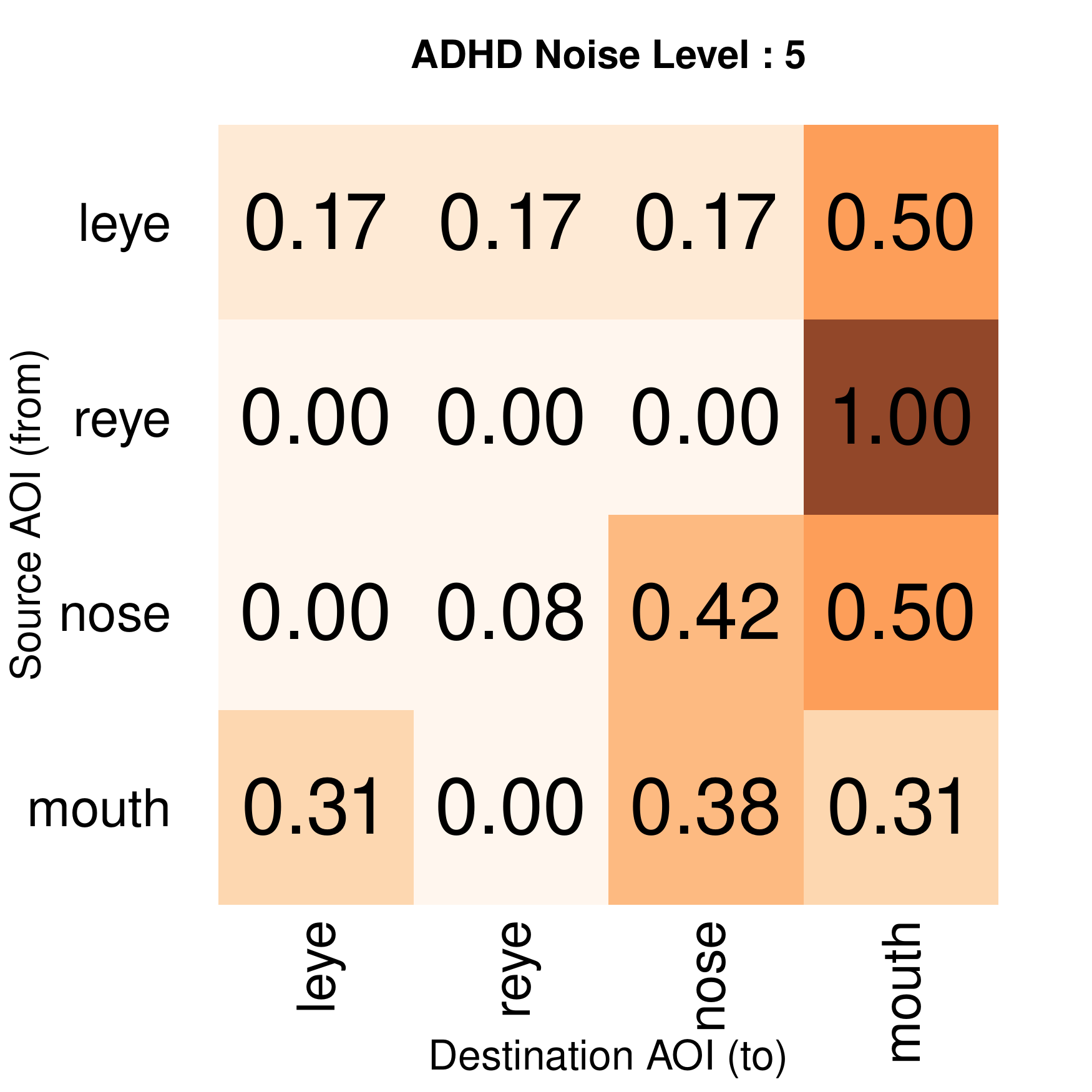}
	}
	\quad
	\subfloat[ADHD:0 dB]{%
		\label{fig:adhd_0}
		\includegraphics[width=.14\textwidth,trim={0 0 0 50},clip]{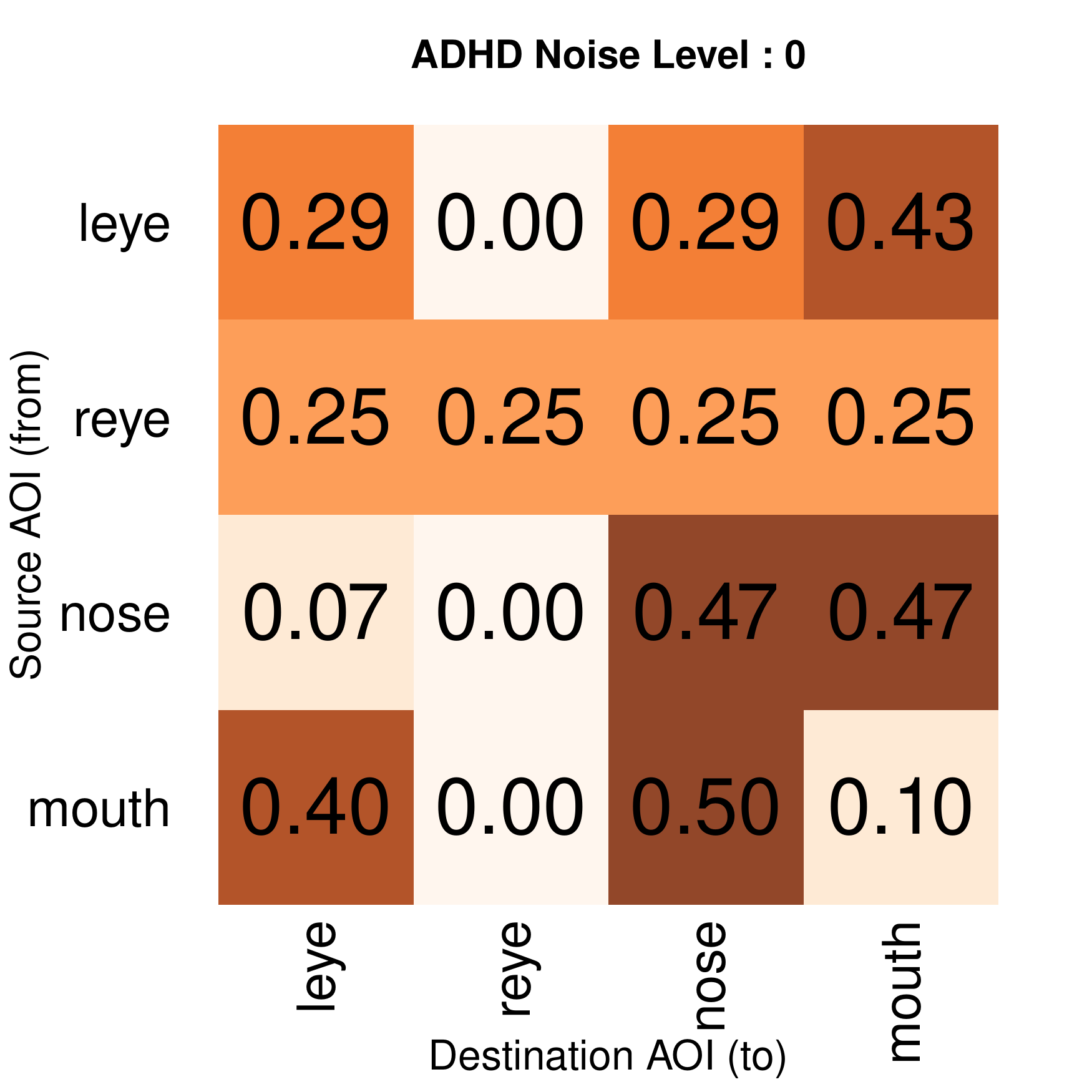}
	}
	\quad

	\subfloat[NT:25 dB]{%
		\label{fig:nonadhd_25}
		\includegraphics[width=.14\textwidth,trim={0 0 0 50},clip]%
		{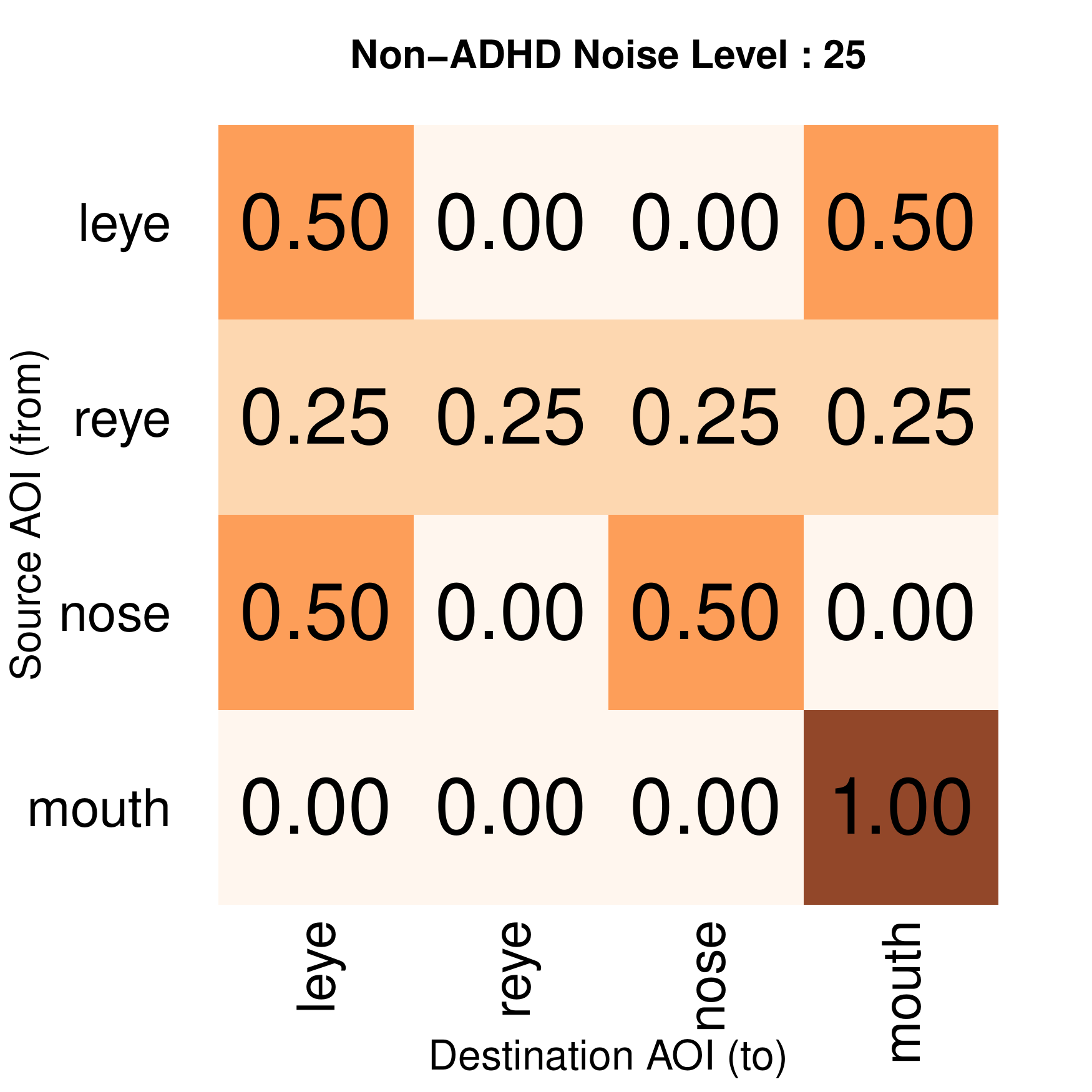}
	}
	\quad
	\subfloat[NT:20 dB]{%
		\label{fig:nonadhd_20}
		\includegraphics[width=.14\textwidth,trim={0 0 0 50},clip]%
		{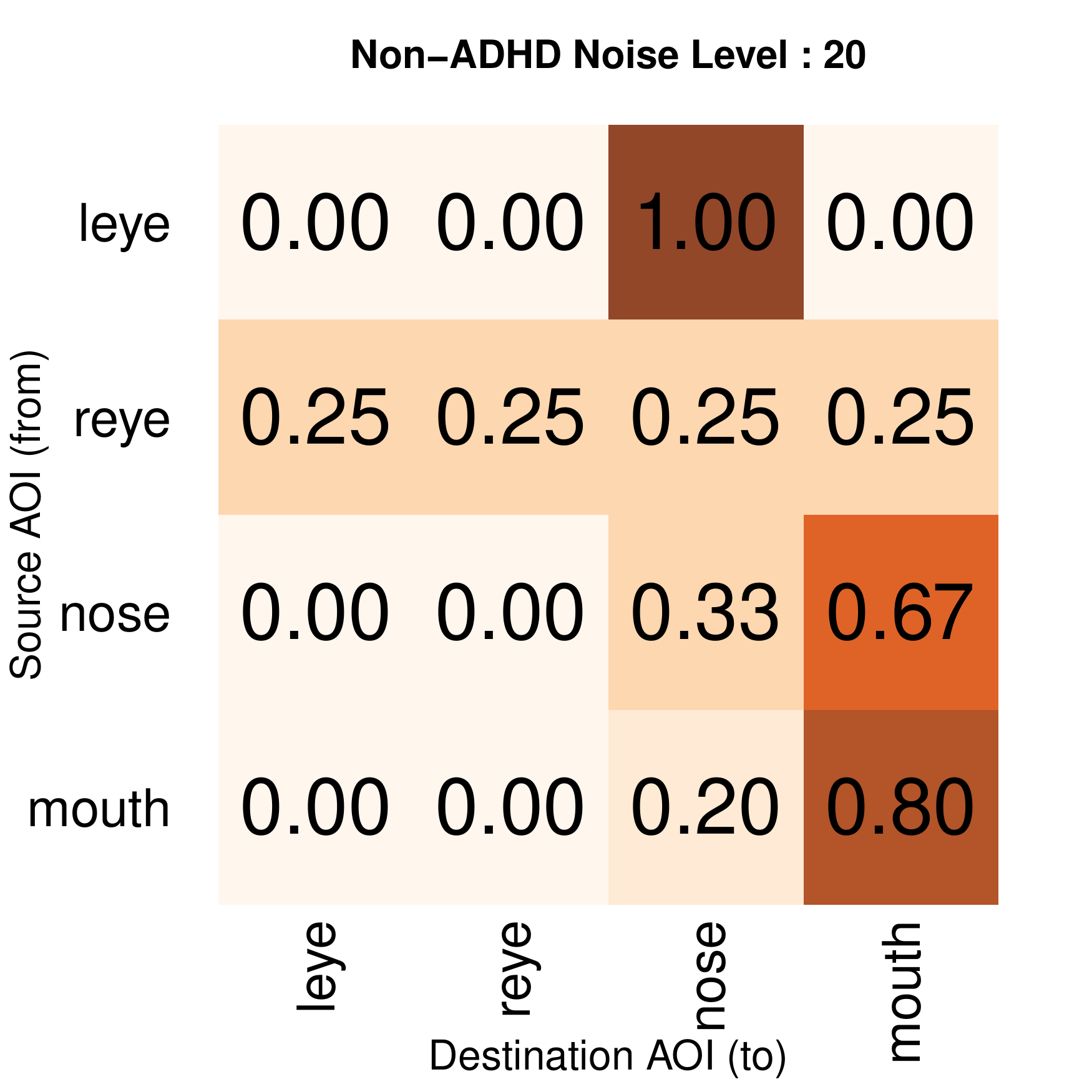}
	}
	\quad
	\subfloat[NT:15 dB]{%
		\label{fig:nonadhd_15}
		\includegraphics[width=.14\textwidth,trim={0 0 0 50},clip]%
		{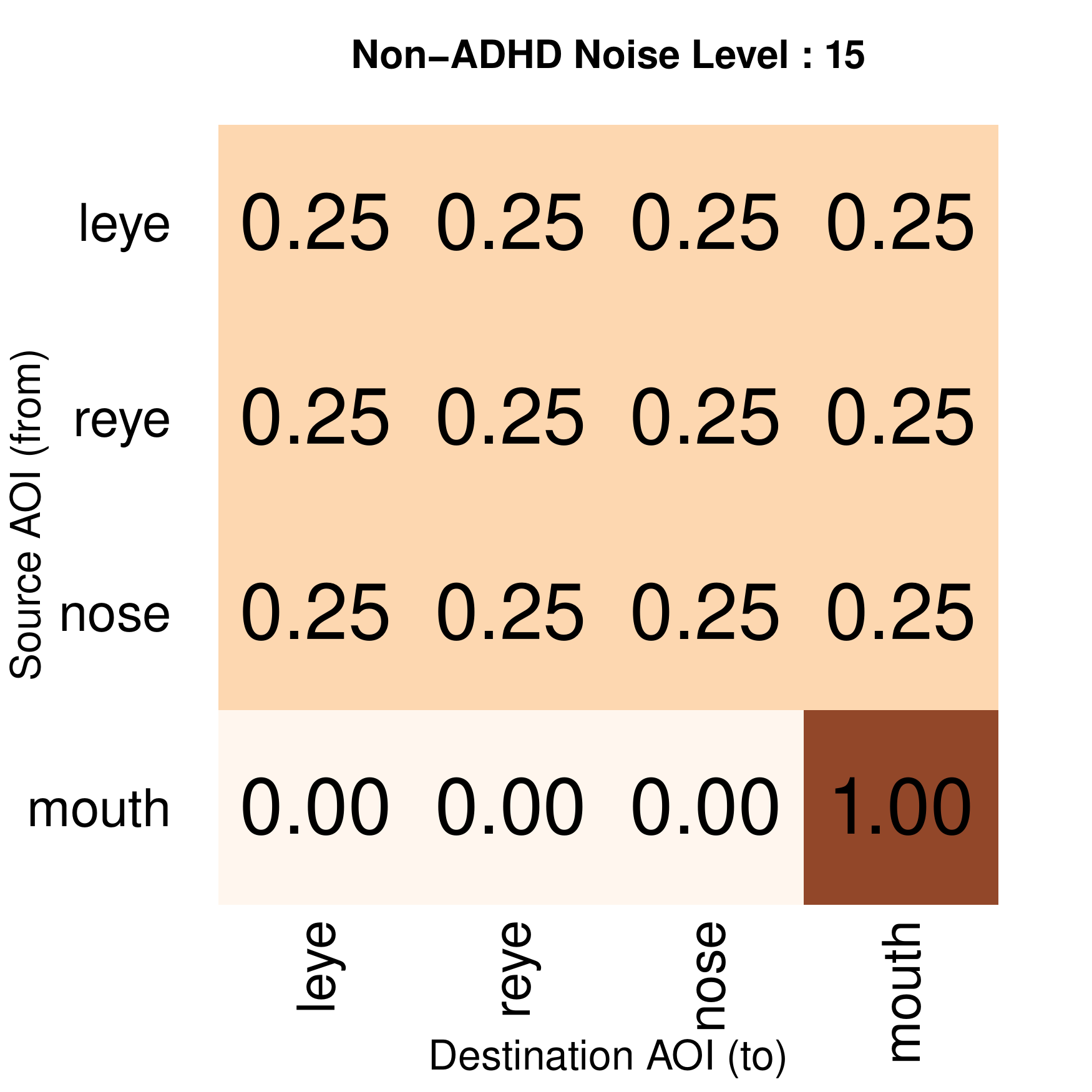}
	}
	\quad
	\subfloat[NT:10 dB]{%
		\label{fig:nonadhd_10}
		\includegraphics[width=.14\textwidth,trim={0 0 0 50},clip]%
		{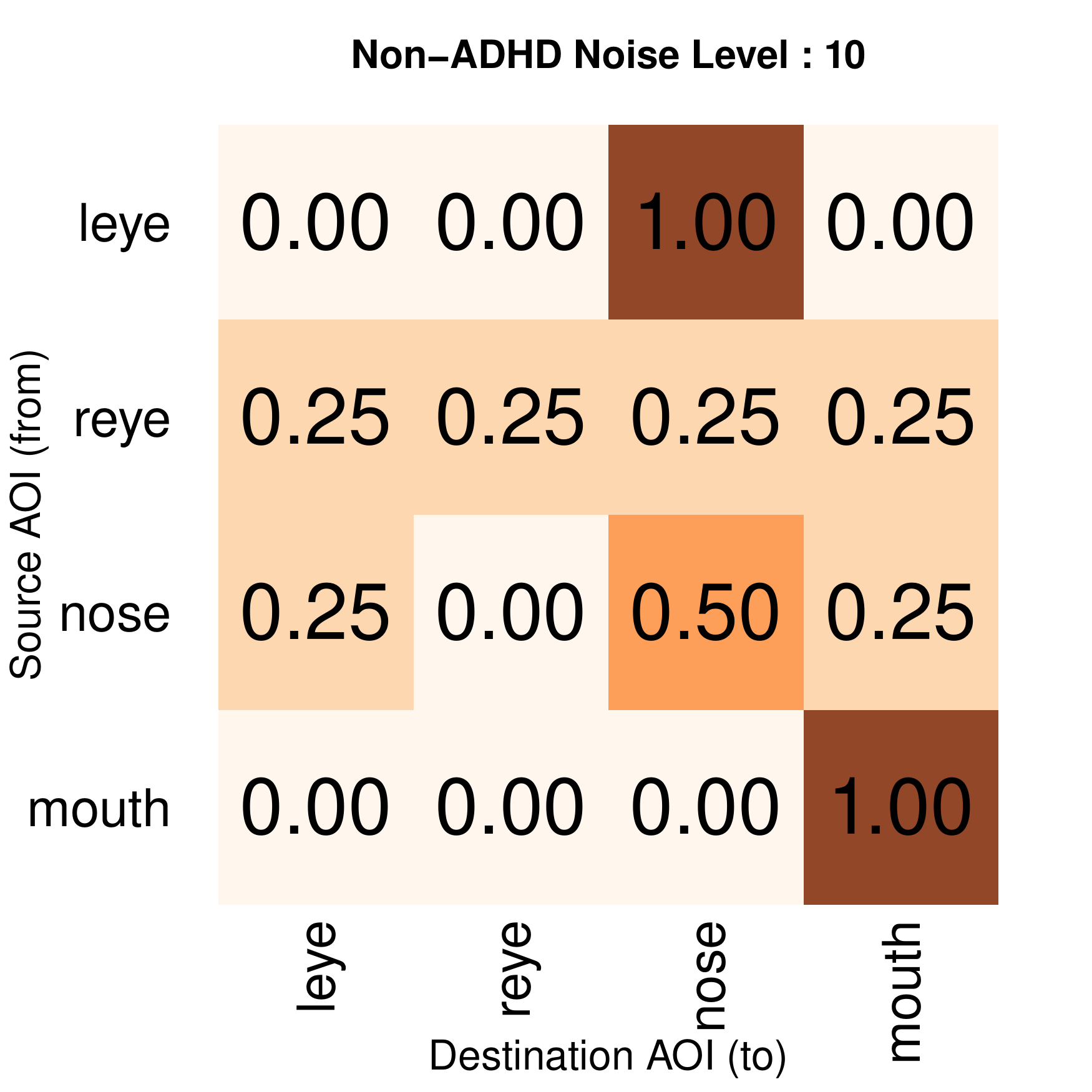}
	}
	\quad
	\subfloat[NT:5 dB]{%
		\label{fig:nonadhd_5}
		\includegraphics[width=.14\textwidth,trim={0 0 0 50},clip]{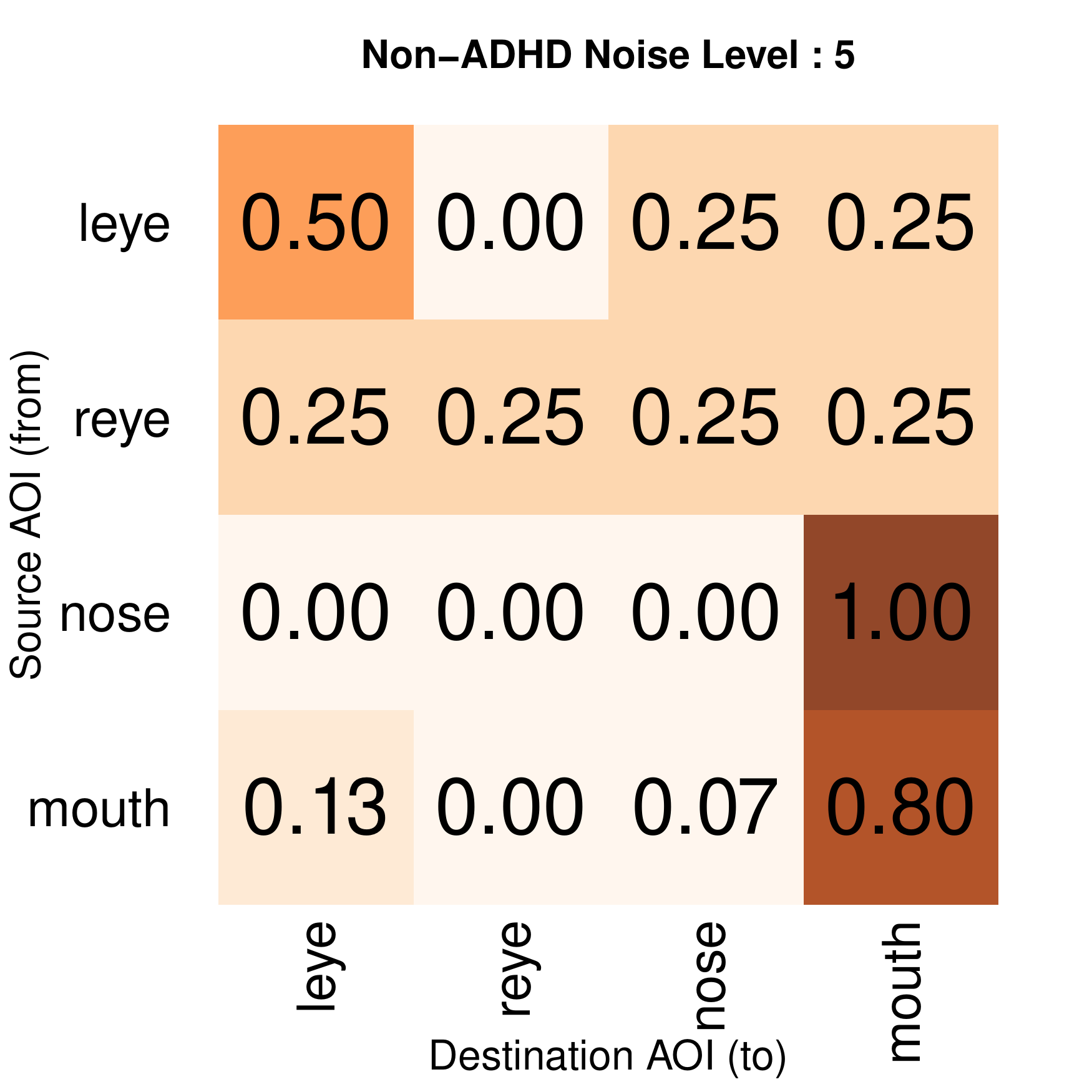}
	}
	\quad
	\subfloat[NT:0 dB]{%
		\label{fig:nonadhd_0}
		\includegraphics[width=.14\textwidth,trim={0 0 0 50},clip]{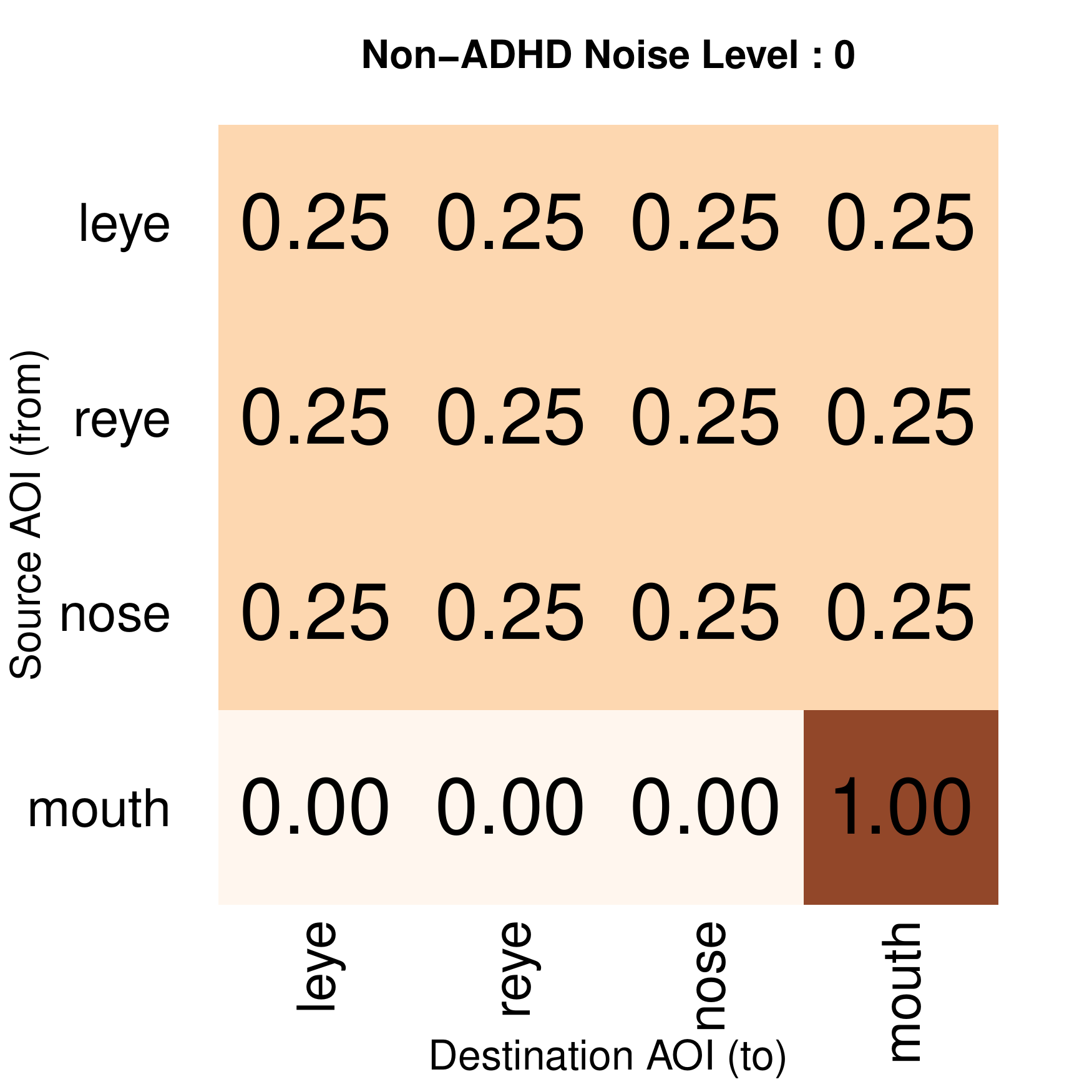}
	}
	\caption{Gaze transition matrices of ADHD and NT participants,
	\textmd{at varying levels of background noise yielding six SNR levels: 0 to 25 dB with 5dB increments.}
		}
	\label{fig:tm}
\end{figure*}

The gaze transition matrices \cite{krejtz2015gaze} indicate the probability of transition of gaze between two AOIs.
Figure \ref{fig:tm} shows the computed gaze transition matrices for ADHD and NT participants at gradually increasing levels of background noise.

Gaze transition matrices for different listening conditions suggest that, in general, participants with ADHD tend to make unpredictable gaze transitions at different difficulty levels of the task whereas NT participants tend to make gaze transition from any AOI to mouth region regardless the difficulty level of the task. 
Interestingly, it can be observed that participants with ADHD tend to re-fixate on the left eye region at 20 dB SNR, where the task is relatively easy.

We calculated the gaze transition entropy to determine the overall distribution of attention over AOIs. 
Small entropy values indicate predictable gaze transitions among AOIs, while large entropy values indicate less predictable gaze transitions among AOIs when transitioning from any source AOI to any destination AOI with similar probabilities \cite{krejtz2015gaze}.

Corresponding transition entropies of computed gaze transition matrices 
are shown in Table \ref{tab:entropy_and_ipa}.
There was no significant main effect of the SNR, $F(1.9,17.18)=2.3, p>0.05$, or the group, $F(1,9)=0.00, p>0.9$ on transition entropies, 
indicating no difference among participant groups or SNRs.
Also, there was no significant interaction effect between SNR and participant group, $F(1.9,17.18)=0.964, p>0.3$.
Table \ref{tab:entropy_and_ipa} shows a tendency of higher entropy for both ADHD and NT participants during the most difficult listening condition (0 dB), indicating less predictability in gaze transitions.
Also, $t$-tests on transition entropies of participants at each SNR  (i.e. without aggregating per participant) showed a significant effect
for the NT group, 
at 0 dB compared to the other listening conditions (all $p<0.03$).

\subsection{The index of pupillary activity (IPA)}

The IPA is calculated using a wavelet-based algorithm that relies on wavelet decomposition of the pupil diameter signal, and its wavelet analysis.
For the IPA calculation, we used Daubechies-4 wavelet for a 60 Hz signal as suggested in \cite{duchowski2018index}.
Low IPA counts reflect little cognitive load whereas high IPA counts indicate strong cognitive load \cite{duchowski2018index}.

\begin{table}[hbt!]
    \centering
    \caption{Gaze Transition Entropy and IPA of ADHD and NT Participants.
    }
    \label{tab:entropy_and_ipa}
    \begin{tabular}{|r|c|c|c|c|}
        \hline
        \multirow{2}{*}{SNR} & \multicolumn{2}{c|}{Entropy} & \multicolumn{2}{c|}{IPA} \\
        \cline{2-5}
        & ADHD          & NT            & ADHD          & NT \\
        \hline
        25 dB    & $0.53\pm0.02$ & $0.59\pm0.02$ & $0.29\pm0.03$ & $0.36\pm0.02$ \\
        
        20 dB    & $0.59\pm0.02$ & $0.56\pm0.02$ & $0.29\pm0.02$ & $0.39\pm0.02$ \\
        
        15 dB    & $0.59\pm0.05$ & $0.60\pm0.04$ & $0.33\pm0.02$ & $0.31\pm0.02$ \\
        
        10 dB    & $0.62\pm0.04$ & $0.58\pm0.04$ & $0.35\pm0.03$ & $0.34\pm0.03$ \\
        
        5 dB     & $0.61\pm0.05$ & $0.56\pm0.04$ & $0.36\pm0.02$ & $0.33\pm0.02$ \\
        
        0 dB     & $0.66\pm0.07$  & $0.69\pm0.06$  & $0.30\pm0.02$ & $0.30\pm0.02$ \\
        \hline
    \end{tabular}
    \vspace{-2pt}
\end{table}

There was no significant main effect of the SNR, $F(5,45)=1.371, p>0.05$, or of 
the group, $F(1,9)=30.7, p>0.05$ on IPA counts, indicating that cognitive load did not differ among participant groups or SNRs.
We observed a significant interaction effect between SNR and group, $F(5,45)=3.265, p<0.02$ on IPA counts such that cognitive load on different SNRs differed in ADHD and NT groups.
Contrasts revealed significant interactions when comparing SNRs 25 dB and 0 dB,$F(1,9)=5.26, p<0.05$, and SNRs 20 dB and 0 dB, $F(1,9)=7.27, p<0.03$.
These effects reflect that the cognitive load differed significantly among hardest and easiest listening conditions between the two groups
(see Table \ref{tab:entropy_and_ipa}).
The remaining contrasts revealed no significant interaction when comparing two groups to different listening conditions, $p>0.05$.

At the hardest listening condition, we expect listening demands to be greater for participants with ADHD, yielding a significant difference in IPA counts, because their innate WMC is lower compared to the NT participants \cite{banich2009neural, alderson2013attention,michalek2014effects}, thus perform significantly different on QuickSIN task.
But, $t$-tests on IPA counts of participants at sentence level (i.e. without aggregating per participant) yielded no significant difference between the ADHD and NT groups, $p>0.08$ for all SNRs, except 15 dB SNR. 
Interestingly, at 15 dB SNR, IPA counts of participants at sentence level yielded a significant difference between the ADHD and NT groups, $p<0.04$ indicating a significant difference in cognitive load between the two groups.
Since cognitive load inherently reduces WMC \cite{chandler1991cognitive,sweller1990cognitive}, we expected participants with ADHD to do worse at 15 dB SNR, as their cognitive load is high and their WMC does not commensurate with NT participants. 
The expected behavior is confirmed by the significant performance difference observed in the evaluation of the number of keywords recalled between ADHD and NT groups at 15 dB SNR.

\section{Discussion}

Our results indicate ADHD, and NT adolescents perform equally likely in the SIN task where audiovisual cues are present when the task difficulty is very high or very low.
However, significant differences of QuickSIN performance between 
participants groups 
were observed at 15 dB SNR where ADHD and NT participants recalled 4.7 and 4.9 keywords on average respectively.
Fixation counts on AOIs suggested that  both groups had a strong preference to look at the mouth of the speaker at the easiest listening condition whereas, both groups looked at the nose of the speaker at the hardest listening condition.
In all other listening conditions, NT participants preferred to look at the mouth 
whereas participants with ADHD mostly looked at the left eye.
The NT individuals perceive audiovisual cues more accurately from the RVF \cite{kimura1973asymmetry}, whereas ADHD orient attention to LVF targets due to the lateralized deficit in visual-spatial attention \cite{carter1995asymmetrical,voeller1988attention,mitchell1990reaction,heilman1991possible}.
Our eye tracking observations align with literature, confirming that young adults with ADHD orient attention to LVF when perceiving audiovisual cues.
Gaze transition matrices and transition entropies show that participants with ADHD tend to make unpredictable gaze transitions at different difficulty levels of the task whereas NT participants tend to re-fixate on the mouth region.

Interestingly, only at 15 dB SNR, IPA counts yielded a significant difference between the two groups, where QuickSIN performance also yielded a significant difference between them. 
This indicates that 15 dB SNR is when the noise shifts to a point where processing of speech becomes less automatic and relies more on increased cognitive load. 
These findings are consistent with the possibility that audiovisual cues, in general, are processed in such a way that WMC or cognitive load are not consistently impacted in increasing levels of background noise for NT group \cite{michalek2018independence}.

\section{Conclusion}

Our work presents an analysis of audiovisual SIN performance for young adults with ADHD compared to age-matched controls using eye-tracking measures. We analyzed the performance of the participants and eye-movement parameters such as fixation count on AOIs, gaze transition entropy, and IPA. 
We observed that participants with ADHD primarily fixated on the left eye of the speaker whereas NT group fixated on the right eye,
supporting the literature that ADHD orients attention to the LVF whereas NT individuals orient attention to the RVF.
When the task difficulty was at a medium level with 15 dB SNR, we observed a significant difference in cognitive load as well as performance between the two groups.

In the future we expect to explore eye movement behavior when scanning the speaker's face in terms of advanced eye movement metrics such as coefficient $\kappa$ of measurement of focal or ambient viewing
, and a larger representation of ADHD and NT adolescents. 

\balance
\bibliographystyle{ACM-Reference-Format}
\bibliography{reference}

\end{document}